\documentclass[times, twocolumn]{aastex631}
\usepackage{graphicx}
\usepackage{amssymb}
\usepackage{mathrsfs} 
\usepackage{natbib}
\usepackage{amsmath}
\usepackage[english]{babel}
\usepackage{url}
\usepackage{hyperref}
\usepackage{xcolor}
\usepackage{xspace}
\usepackage{CJK}

\newcommand{\haffet}{{\tt HAFFET} \xspace}
\newcommand{\superbol}{{\tt SuperBol}\xspace}
\newcommand{\emcee}{{\tt emcee}\xspace}

\begin{document}
\begin{CJK*}{UTF8}{gbsn}
\title{Magnetar Flare-Driven Bumpy Declining Light Curves in Hydrogen-poor Superluminous Supernovae}

\author[0000-0001-7950-0531]{Xiao-Fei Dong (董晓菲)}
\affiliation{Institute for Frontier in Astronomy and Astrophysics, Beijing Normal University, Beijing 100875, People's Republic of China}
\affiliation{Department of Astronomy, Beijing Normal University,
Beijing 100875, People's Republic of China}

\author[0000-0002-8708-0597]{Liang-Duan Liu (刘良端)}
\affiliation{Institute of Astrophysics, Central China Normal University, Wuhan 430079, China;}
\affiliation{Key Laboratory of Quark and Lepton Physics (Central China Normal University), Ministry of Education, Wuhan 430079, China}

\author[0000-0002-3100-6558]{He Gao (高鹤)}
\affiliation{Institute for Frontier in Astronomy and Astrophysics, Beijing Normal University, Beijing 100875, People's Republic of China}
\affiliation{Department of Astronomy, Beijing Normal University,
Beijing 100875, People's Republic of China}

\author[0000-0002-2898-6532]{Sheng Yang (杨圣)}
\affiliation{Henan Academy of Sciences, Zhengzhou 450046, Henan, China}
\affiliation{Department of Astronomy, The Oskar Klein Center, Stockholm University, AlbaNova, 10691 Stockholm, Sweden}

\correspondingauthor{He Gao}
\email{gaohe@bnu.edu.cn}

\correspondingauthor{Liang-Duan Liu}
\email{liuld@ccnu.edu.cn}

\begin{abstract}

Recent observations indicate that hydrogen-poor superluminous supernovae often display bumpy declining light curves. However, the cause of these undulations remains unclear. In this paper, we have improved the magnetar model, which includes flare activities. We present a systematic analysis of a well-observed SLSNe-I sample with bumpy light curves in the late-phase. These SLSNe-I were identified from multiple transient surveys, such as the Pan-STARRS1 Medium Deep Survey (PS1 MDS) and the Zwicky Transient Facility (ZTF). Our study provides a set of magnetar-powered model light curve fits for five SLSNe-I, which accurately reproduce observed light curves using reasonable physical parameters. By extracting essential characteristics of both explosions and central engines, these fits provide valuable insights into investigating their potential association with gamma ray burst engines. We found that the SLSN flares tend to be the dim and long extension of the GRB flares in the peak luminosity versus peak time plane. Conducting large-scale, high cadence surveys in the near future could enhance our comprehension of both SLSN undulation properties and their potential relationship with GRBs by modeling their light curve characteristics.

\end{abstract}


\section{Introduction} \label{sec:intro}
Superluminous supernovae (SLSNe) have been a hot topic in supernova research for nearly two decades since the first discovered events \citep[e.g., SN 2005ap;][]{Quimby2007}. SLSNe are characterized by their high luminosity, which can be up to tens to hundreds of times brighter than typical supernovae \citep[see][for a review]{Gal19}.  SLSNe can be divided into two categories based on their spectra: hydrogen-rich (SLSN-II) and hydrogen-poor (SLSN-I). It is widely believed that the high luminosity of SLSN-II is primarily attributed to the interaction between the supernova ejecta  with their massive hydrogen-rich circumstellar medium (CSM) \citep{Smi07,Che11}.

The total energy radiated by a typical SLSN-I is approximately a few times $10^{51}$ ergs, if all of this energy is provided by the decay of $^{56}$Ni, it would require at least several solar masses of $^{56}$Ni. However, it is very difficult for a regular supernova nucleosynthesis process to produce so much $^{56}$Ni \citep{Umeda2008}. The conventional model of radioactive power is facing a significant challenge due to the exceptionally high luminosity exhibited by SLSN-I\footnote{LCs of some slowly evolving SLSNe could be explained by radioactive decay of massive amounts of $^{56}$Ni synthesized in a pair-instability supernova explosion \citep[e.g.,][]{Gal2009,guti22}.  However, no SLSN has been convincingly shown to be consistent with a radioactive-powered model. Even those that have a similar decay rate as $^{56}$Co decay are too blue and rise too quickly. The vast majority of rapidly evolving SLSNe-I are unlikely to be primarily powered by radioactive decay.}. There are currently two main scenarios for the energy source of SLSN-I, namely the CSM interaction scenario  \citep{Che11,Gin12,Cha12} and the magnetar-powered scenario \citep{Kas10,Woo10}. The former scenario is similar to SLSN-II, except that the SN is surrounded by a massive hydrogen-poor CSM. The latter scenario considers energy injection from the spin-down of a newborn rapidly rotating magnetar.  Numerous prior studies have demonstrated that these two scenarios are capable of explaining the general shape of the photometric light curves (LCs) of SLSNe-I \citep[e.g.,][]{Cha13,Ins13,Wang15,Pra17,Yu17,Liu17,Nic17,Chen2023b}, which makes it difficult to distinguish one scenario from the other. Further photometric observations are necessary to discriminate between the energy source models.

Currently, time-domain surveys are gathering substantial amounts of well-observed SLSNe-I. The magnetar-powered model or CSM interaction alone cannot explain undulations and occasional significant post-peak bumps observed in some SLSNe-I LCs \citep[e.g.,][]{Nicholl_bn,Inserra2017Interaction,Yan2020Ib,Gomez2021,Fiore2017gci,SN2020qlb,guti22,SN2020wntdip,Zhu2023}. Recently, systematic studies of the prevalence and properties of post-peak bumps of SLSNe have found that these fluctuation structures may be relatively common \citep{HosseinzadehSLSNstat,Chen2023b}.  

The identification of post-peak bumps in certain SLSNe-I may provide novel understanding into the energy source mechanisms of these SNe.  \cite{Liu18} has shown that a multiple ejecta-CSM interaction model can explain the undulating bolometric LC of iPTF 15esb and iPTF 13dcc. The absence of relatively narrow emission lines in most spectral readings of SLSNe-I is often seen as a major challenge to the CSM-interaction as the primary power source. 

In the magnetar-powered model, the LCs of SLSNe are determined by the behavior of the central engine over time. If the spin-down of the magnetar is mainly through magnetic dipole radiation, then a relatively smooth light curve can be expected. In fact, magnetars have intermittent and violent energy burst activities, which is an important observational feature of magnetars in the Milky Way \citep[see][for a review]{Kaspi2017}. Furthermore, in about half of \emph{Swift} gamma-ray bursts (GRBs) X-ray afterglows, a large number of rapidly rising and falling X-ray flares have been discovered \citep{Burrows2005,Falcone2007,Wang2013,yi2016}. This strongly suggests that the GRB engines have undergone numerous intermittent and energetic activities. It is proposed that a millisecond spin-down magnetar could simultaneously power long GRBs and SLSNe \citep[e.g.,][]{Margalit2018}. They both have a similar host preference in metal-poor dwarf galaxies with 
relatively high star formation rate \citep[e.g.,][]{Lunnan2014,ChenTW2015,Perley2016,ChenTW2017,Schu18}. Besides, the spectra of some SLSNe-I in their nebular phase are remarkably similar to those of GRB-SNe \citep[e.g.,][]{Nic16b,Jerk17}. 
Given the significant correlation between GRBs and SLSNe \citep[as note by][]{Yu17,Metzger2018,Liu2022}, it is reasonable to assume that flare activity may also be caused by the central magnetar of SLSNe. These intermittent flare activities may reheat the SNe ejecta in later stages, and then show undulations characteristics on the LC \citep{Yu&Li2017}.

This paper focuses on the origin of late-time undulations and whether the magnetar model with flare components can replicate the observed bumpy declining light curves (LCs) of SLSNe-I. To achieve this, we have updated the magnetar model by incorporating the Markov chain Monte Carlo (MCMC) technique and conducted a systematic analysis using well-sampled late-phase LCs of SLSNe-I, including multi-color LC fits and parameter comparisons. The paper is structured as follows: Section \ref{sec:data} describes a sample of well-observed SLSNe-I with post-peak bumps; Section \ref{sec:model} provides a brief description of our model and fitting process for the LCs; Section \ref{sec:results} presents statistics on fitting parameters and compares them to GRB flares; finally, in Section \ref{sec:discussion}, we draw conclusions from our findings and discuss their implications.

\section{Sample Selection} \label{sec:data}

Previously, \citet{HosseinzadehSLSNstat} found that the majority (44-76\%) of their SLSNe-I sample possess late-time LC fluctuations. Similar undulations were detected in 34\%-62\% SLSNe-I discovered in ZTF Phase-I Survey \citep{Chen2023b}. However, these work mostly focused on observational statistics, whereas the existing theoretical study of LC fluctuations mostly focused on individual objects \citep[e.g.,][]{Fiore2017gci,chugai2022,SN2020wntdip,SN2020qlb}. Here, we  perform systematic fits on SLSNe-I LCs with bumpy features under the magnetar-powered model. To investigate the undulations of SLSNe-I, we need adequate observational data on their post-peak bumps. We collected publicly available data on SLSNe-I in the Open Astronomy Catalog (OAC)\footnote{https://github.com/astrocatalogs/OACAPI} and the Bright Transient Survey explorer \citep[BTS,][]{fremling2020,perley2020} catalog\footnote{https://sites.astro.caltech.edu/ztf/bts/explorer.php}. Our sample includes all SLSNe-I that meet the criteria listed in Table \ref{tab:Criteria}. Specifically, We performed the selection with the following three steps:

\begin{enumerate}
\item This object has been spectroscopically classified as a SLSN-I.

\item The light curve of SLSN must be well observed with a distinct primary peak in at least two filters, accompanied by at least five data points for both the rising and declining phases. 

\item The SLSN late-time LCs deviate from the smooth decline trend predicted by the magnetar spin-down model and feature in one or more additional post-peak ``bumps'', which contain a complete rising and declining structure (i.e., an obvious second or third peak) and should be identified with residuals larger than $3\sigma$. 
\end{enumerate}

\begin{table}[ht!]
\caption{Criteria for selecting SLSNe-I with late-time bumps.}
\begin{center}
\begin{tabular}{ccc} \hline\hline
Step & Critiria & Sample Size \\ 
\hline 
1 & Spectroscopically identified as SLSNe-I & 182\\
2 & Main peak well sampled \& & 80 \\
& no large LC gaps &  \\
3 & $\textgreater 3\sigma$ deviation from the magnetar model \& &5 \\
&  complete second/third peak(s)& \\
\hline
\end{tabular}
\end{center}
\label{tab:Criteria}
\end{table}

In addition, SLSNe-I with large gaps or deficient rising phase in their LCs (e.g., SN 2017gci) are excluded from our sample. Following the criteria, we finally selected 5 SLSNe-I\footnote{SN 2017egm is a nearby SLSN-I with undulated LCs that has attracted a lot of attention. Its main peak displays a typical inverted triangle shape \citep[e.g.,][]{Bose18,Tsv22,Zhu2023}, which is challenging to explain using the magnetar model. \citet{Lin23} has analyzed the overall LC evolution of SN2017egm and found that its complex LC can be well modeled by successive CSM interactions. In combination with these discussions and our baseline fitting, we excluded this source from our sample.}. Their basic information and references are listed in Table \ref{tab:infor}. For the three ZTF objects of our sample, we obtained their apparent magnitude using the python-based package \haffet \citep{HAFFET}, which can request data from the ZTF Forced Photometry Services \citep{Masci2019} directly. In our sample, three SLSNe only contain observations in two bands. To avoid introducing additional systematic errors, we fit their multi-color LCs instead of a bolometric one, which is also able to unveil the color information. 

\begin{table}[ht!]
\centering
\caption{Basic Information of 5 Selected SLSNe-I}
\begin{tabular}{lccc} \hline\hline
Name & Redshift & Bands Used& Photometry Reference(s) \\ 
\hline 
PS1-12cil & 0.32 & g, r, i, z&\cite{LunnanPS1} \\
SN 2019stc & 0.1178 & g, r, i& \cite{Gomez2021} \\ 
SN 2021mkr & 0.28 & g, r& ZTF forced-photometry \\ 
SN 2018kyt &  0.1080 &g, r&  ZTF forced-photometry \\ 
SN 2019lsq & 0.1295 & g, r& ZTF forced-photometry \\
\hline
\end{tabular}
\label{tab:infor}
\end{table}

\begin{figure*}[ht!]
\centering
\gridline{\fig{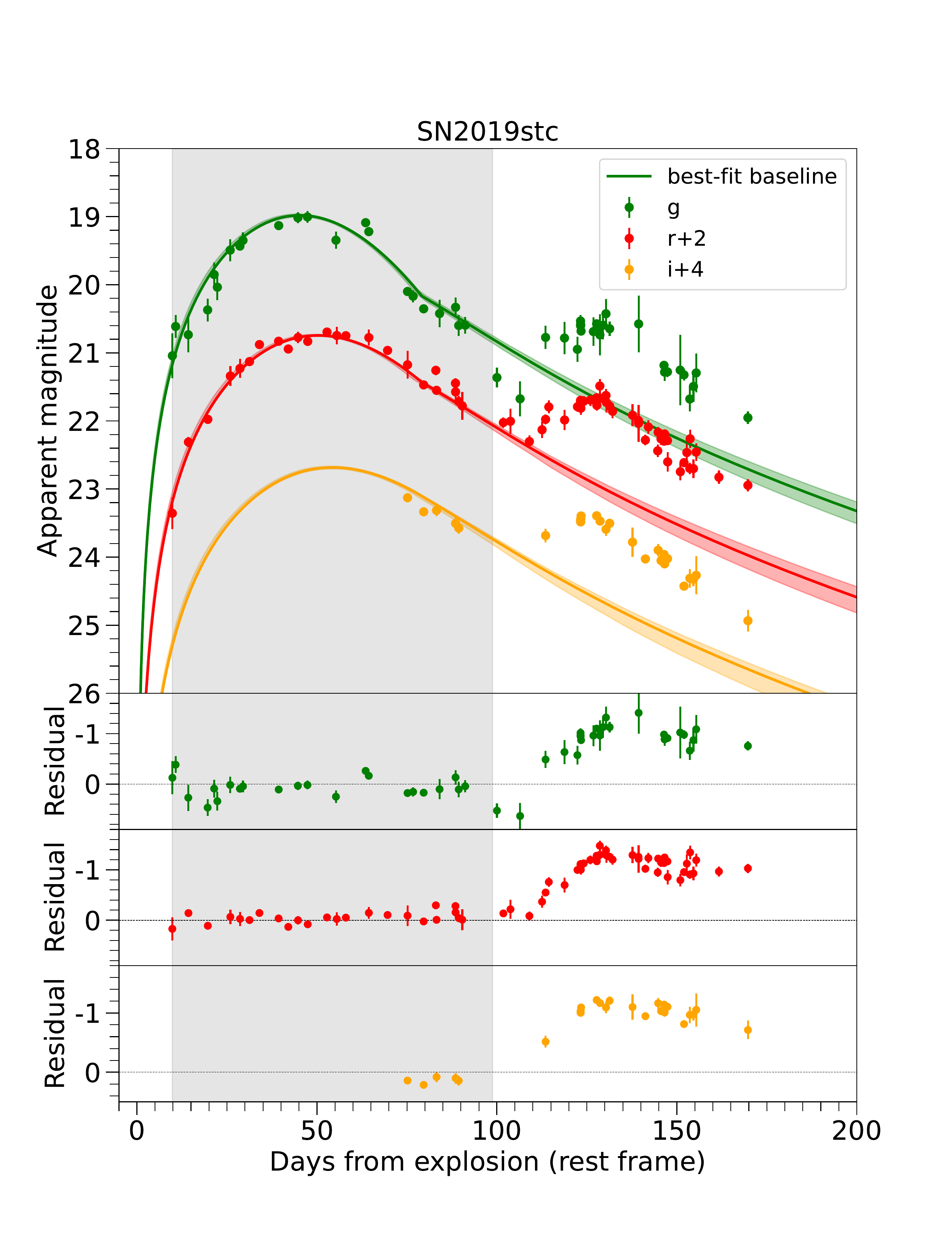}{0.333\textwidth}{(a)}
          \fig{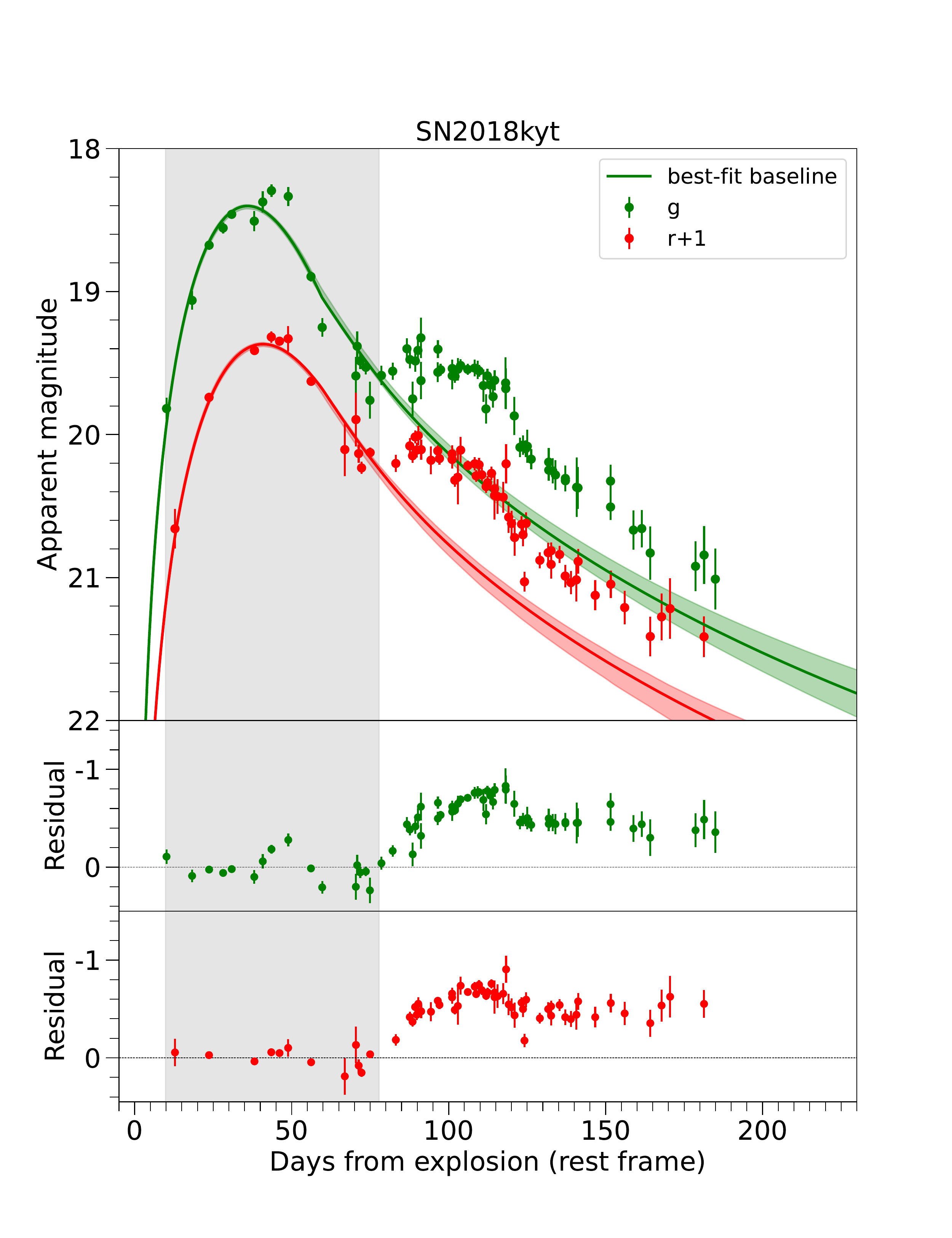}{0.333\textwidth}{(b)}
         \fig{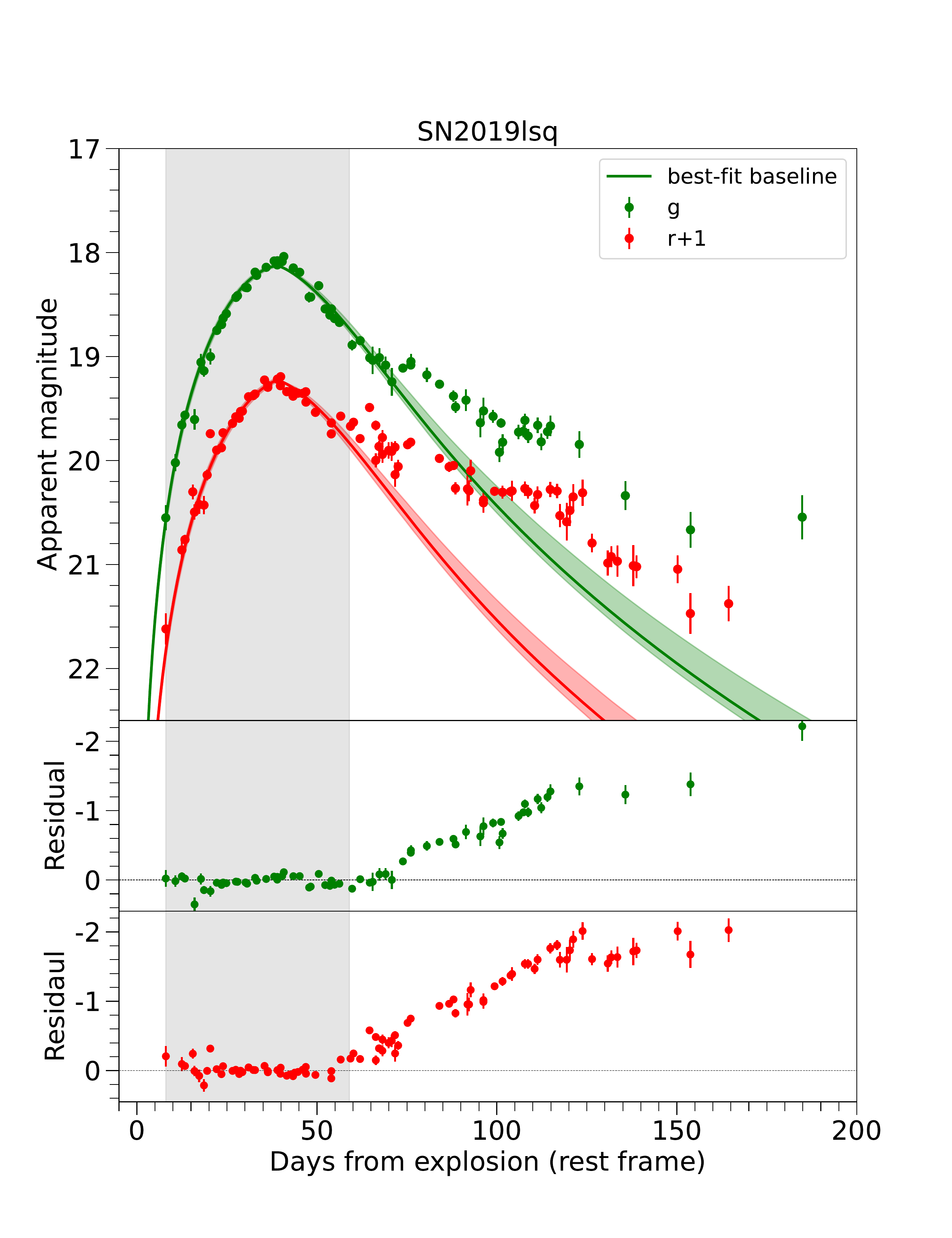}{0.333\textwidth}{(c)}}
\gridline{
          \fig{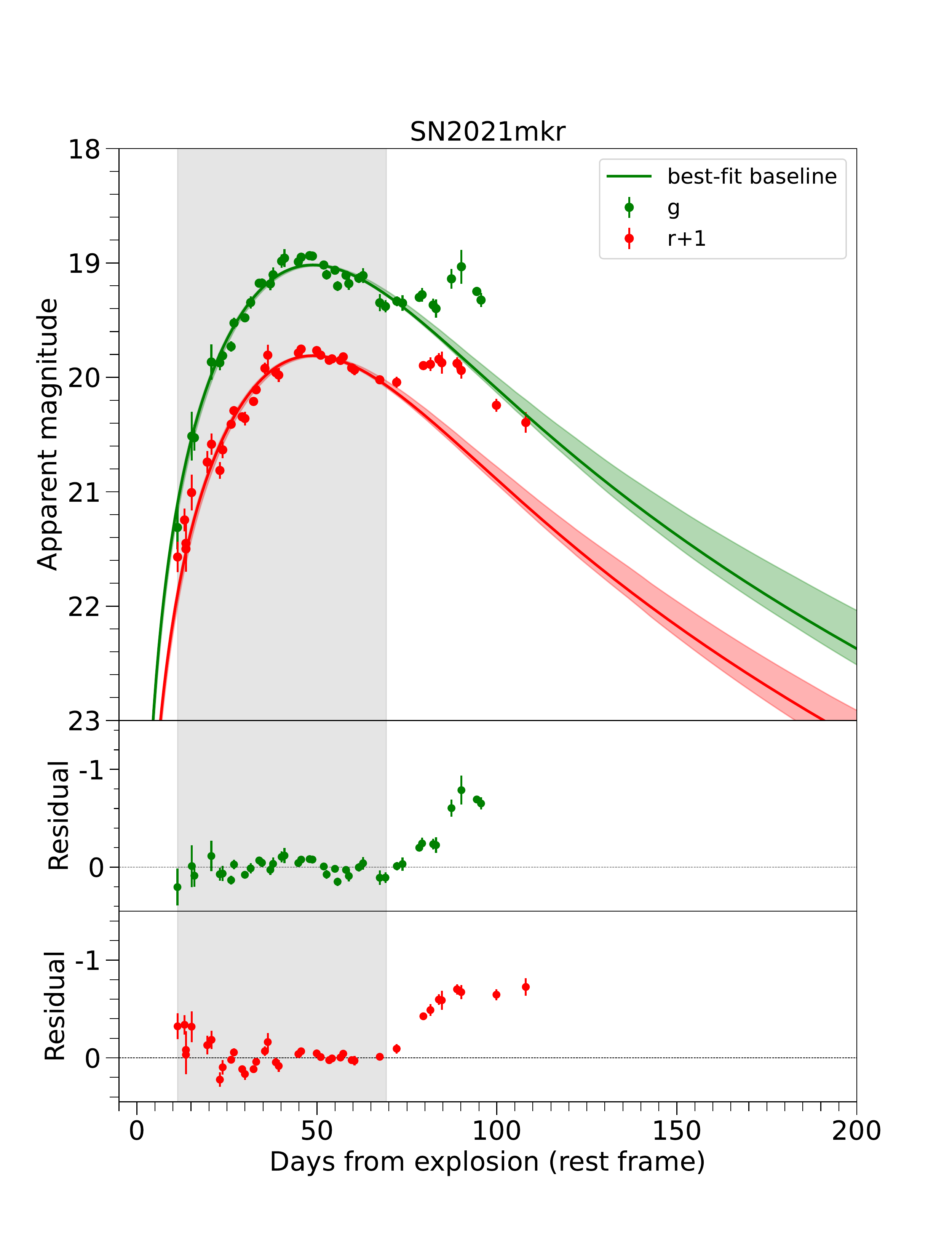}{0.333\textwidth}{(d)}
          \fig{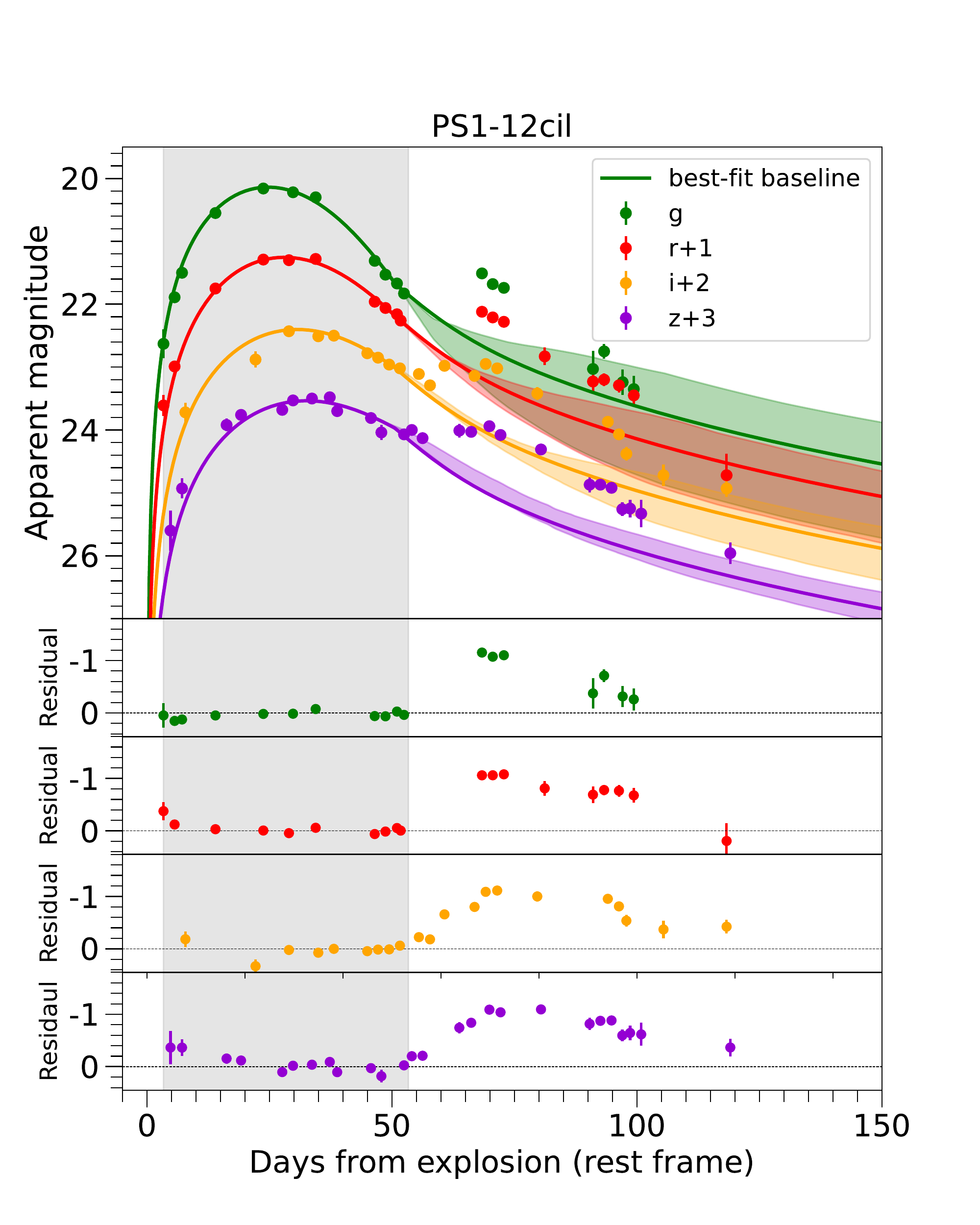}{0.333\textwidth}{(e)}}

\caption{Multi-color LCs compared to their baseline fitting by the spin-down magnetar model. (a) SN 2019stc, (b) SN 2018kyt, (c) SN 2019lsq, (d) SN 2021mkr, (e) PS1-12cil. The green, red, and orange solid lines represent the best baseline fitting results in g, r, and i band, respectively. 2$\sigma$ interval is shown as translucent color range. Residuals with respect to the baseline are shown in the lower panels. Data inside the gray shaded region were used to determine the baseline. The relatively larger error range of PS1-12cil in g and r band is attributed to the large sampling interval. 
\label{fig:baseline}}
\end{figure*}

\section{Light Curves Fitting} \label{sec:model}
Our model assumes that the continuous injection of energy determines the underlying profile of SLSN LCs. The undulations in the LC are caused by flare activities from the central magnetar. Initially, we examine when and how SLSN LCs deviate from the smooth magnetar model. For each SN in our sample, we subtract the best-fitting model from the observed LC to identify characteristics of late-time excess. Subsequently, we will establish a connection between these post-peak bumps and magnetar flare activities.

\subsection{The underlying profile of the SLSNe light curve} \label{sec:baseline}

In the previous magnetar-powered model, the typical assumption is that the spin-down luminosity of the magnetar follows the magnetic dipole radiation as \citep[e.g.,][]{Cha12,Ins13,Nic17}
\begin{equation}
L_{\mathrm{sd}}(t)=L_{\mathrm{sd}, \mathrm{i}}\left(1+\frac{t}{t_{\mathrm{sd}}}\right)^{-2},
\end{equation}
where $L_{\mathrm{sd}, \mathrm{i}}\simeq10^{47} \operatorname{erg~s}^{-1} P_{\mathrm{i},-3}^{-4} B_{\mathrm{p}, 14}^2$ is the initial spin-down luminosity and $t_{\mathrm{sd}} \simeq 2 \times 10^5 \mathrm{~s} P_{\mathrm{i},-3}^2 B_{\mathrm{p}, 14}^{-2}$ is the characteristic spin-down timescale. $P_{\mathrm{i}}$ and $B_{\mathrm{p}}$ are the initial spin period and polar magnetic strength of the magnetar, respectively. Here we take the conventional notation $Q_x=Q/10^x$ in cgs units. This spin-down luminosity of the magnetar inflates a nebula of relativistic particles and radiation inside the cavity evacuated by the expanding supernova ejecta.

Since the energy from the magnetar may be in the form of high-energy photons, we can incorporate the gamma ray leakage effect into the standard supernova diffusion equation to calculate the bolometric luminosity as \citep{Arnett1982}

\begin{equation} \label{Eq:L_bol}
\begin{split}
L_{\rm bol}(t)=e^{-\left(\frac{t}{t_{\rm diff}}\right)^2}\left(1-e^{-At^{-2}}\right)\times \\
\int^t_0 2L_{\rm sd}(t')\frac{t'}{t_{\rm diff}} e^{\left(\frac{t'}{t_{\rm diff}}\right)^2}\frac{dt'}{t_{\rm diff}},
\end{split}
\end{equation}
where the diffusion time $t_{\rm diff}=({2\kappa M_{\rm ej}}/{\beta cv_{\rm ej}})^{\frac{1}{2}}$, determined by ejecta mass $M_{\rm ej}$, velocity $v_{\rm ej}$, and optical opacity $\kappa$. $\beta$ is a constant integration $\sim 13.8$ and $c$ is the speed of light. The factor $\left(1-e^{-At^{-2}}\right)$ accounts for high-energy photons, and $A=3\kappa_{\gamma} M_{\rm ej}/4\pi v_{\rm ej}^2$ \citep{Wang15}, $\kappa_{\gamma}$ is the opacity of gamma rays.

To calculate the multi-band LC of a SLSN, we assume a photospheric temperature as \citep{Nic17}

\begin{equation}
T_{\mathrm{ph}}(t)=\max \left[\left(\frac{L_{\mathrm{bol}}}{4 \pi \sigma_{\mathrm{SB}} v_{\mathrm{ej}}^2 t^2}\right)^{1 / 4}, T_{\text {floor }}\right]
\end{equation}
where $\sigma_{\mathrm{SB}}$ is the Stefan -Boltzmann constant and $ T_{\text {floor }}$ is the photospheric temperature.

In order to obtain the baseline parameters properly, we only utilized observations of the main peak in their LCs to fit the baseline (marked as the shaded grey region in Figure \ref{fig:baseline}). The model fittings were performed using the MCMC technique based on the python-based package \emcee \citep{emcee}. The typical number of iterations for the MCMC algorithm is 5000. All the free parameters used in this model are listed in Table \ref{tab:def}. The maximum likelihood approach was adopted to obtain the best fitting parameters, which are listed in Table \ref{tab:basepara}. We present the best-fitting baseline with residuals for each SLSN in Figure \ref{fig:baseline}, where the underlying LC profile of each SLSN is well described by the magnetar model. With respect to the basic model, the late-time deviations are rather distinct for all 5 sources, which clearly requires an additional physical process. Moreover, the luminosity excess of each source appears almost simultaneously in different filters, suggesting that the bump is wavelength-independent.

\subsection{Light-curve undulations and magnetar flare activities} \label{sec:flare}

Besides magnetic dipole radiation, another way to release the rotation of a magnetar is through erupting magnetic bubbles due to the wind-up of the seed field \citep{Klu1998,Ruderman2000}. This process naturally creates an unpredictable central engine that is significant in understanding the prompt emission of GRBs and X-ray flares \citep{Dai2006}. In nearly half of all GRBs, bright X-ray flares are detected \citep{Burrows2005,Falcone2007,Wang2013,yi2016}. This strongly suggests that the GRB central magnetar have undergone numerous intermittent and energetic activities.  By considering  the strong correlation between GRBs and SLSNe \citep{Yu17,Metzger2018,Liu2022}, it is reasonable to assume that the central magnetar of SLSNe could also trigger flare activities, leading to undulations in their LCs due to delayed release of flaring energy.

We describe the energy release rate resulting from the central magnetar flare using a formula commonly used to fit GRB X-ray flare LCs \citep{yi2016}
\begin{equation}\label{eq:flare}
L_{\rm flare}(t)=L_{\rm flare,pk}\left[\left(\frac{t}{t_{\rm pk}}\right)^{\alpha_1\omega}+\left(\frac{t}{t_{\rm pk}}\right)^{\alpha_2\omega}\right]^{-1/\omega},
\end{equation} 
where $\alpha_1$, $\alpha_2$ and $\omega$ are the structure parameters reflecting the sharpness and smoothness of the flares, which have relatively small impact on the final LC shapes \citep{Yu&Li2017}. In this work, we adopted general fixed values with $\alpha_{\rm 1}=-2$, $\alpha_{\rm 2}=2.5$ and $\omega=5$. $L_{\rm flare,pk}$ and $t_{\rm pk}$ are free parameters representing the flare energy injection and peak time. However, unlike in the case of GRB flares, the energy and time of flare activities cannot be straightforwardly compared to the observed post-peak bumps of SLSNe-I. The main reason is that the flare energy is likely trapped in the optically thick SN ejecta and re-radiated by diffusion. The energy released from the magnetar flare activities can reheat SN ejecta and then cause the ``bump'' feature appearing in the LCs. By substituting the expression of $L_{\rm flare}(t)+L_{\rm sd}(t)$ into Eq.(\ref{Eq:L_bol}) to replace $L_{\rm sd}(t)$, we can obtain the LC after being affected by the flare activities of central magnetar.

\begin{figure*}[ht!] 
\centering
    \includegraphics[width=1.0\textwidth]{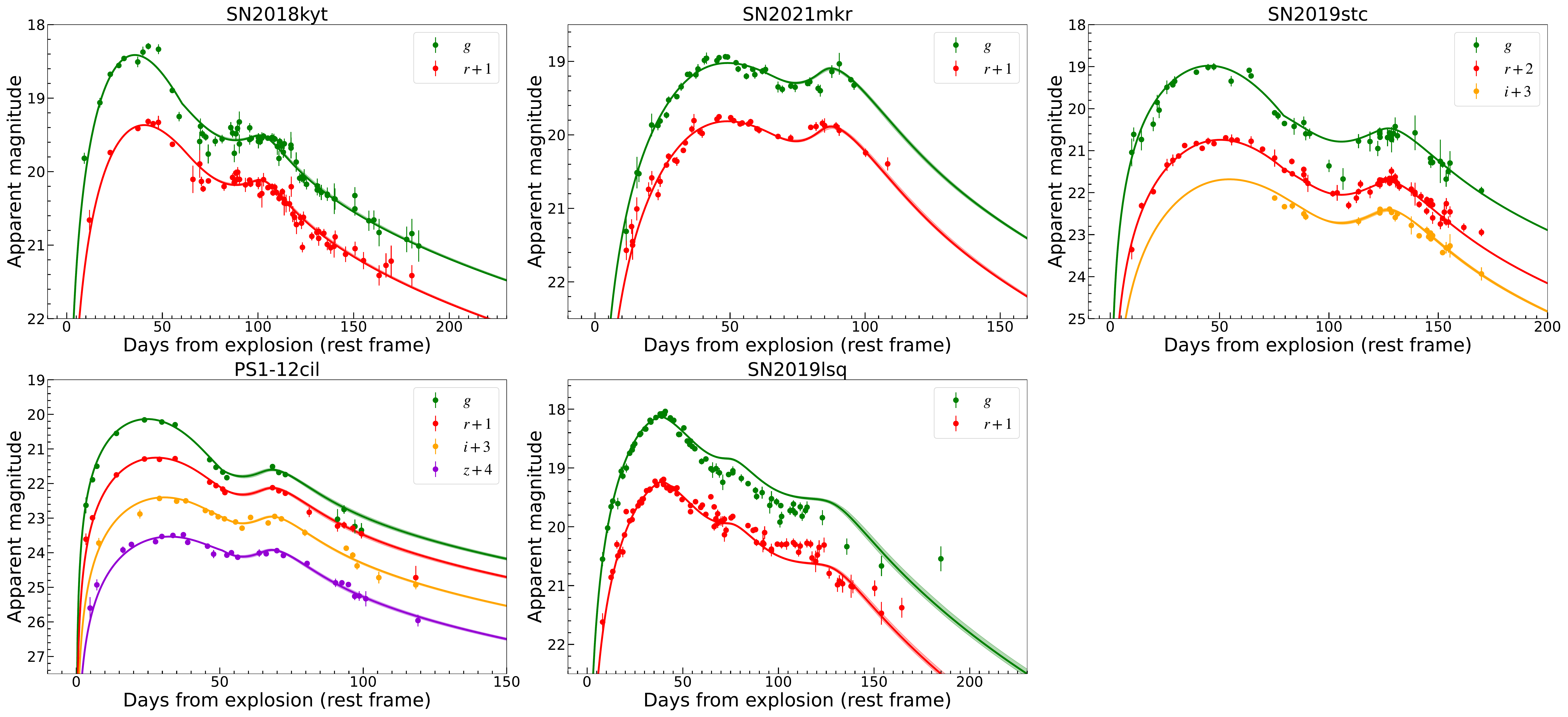}
    \caption{Multi-wavelength observations (dot) with the best magnetar-powered fitting (solid line), where time dilation was corrected. Different colors represent different bands. Data from ZTF force photometry were binned in one day.}
    \label{fig:fit}
\end{figure*}

\begin{figure*}[ht!]
\centering
    \includegraphics[width=1.\textwidth]{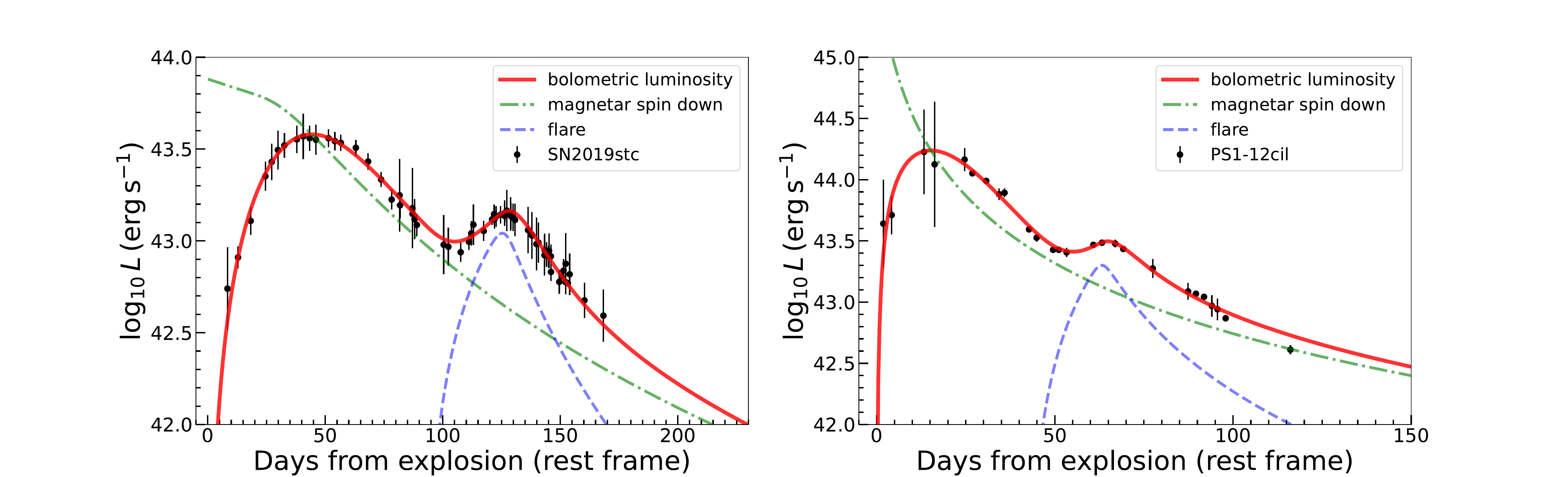}
    \caption{Bolometric LCs (black dot) with magnetar fitting of SN 2019stc and PS1-12cil, where the red solid line represents the bolometric LC, and the green dot line and blue dash line reflect the spin-down magnetar baseline and the flare component.}
    \label{fig:bol}
\end{figure*}

By adding additional flare components to the best-fitting baseline, we successfully reproduced the multi-color LCs of the 5 sources. The final fitting results are presented in Figure \ref{fig:fit}, where two flares are introduced to explain the LCs of SN 2019lsq and one for other objects. The corresponding parameters are listed in Table \ref{tab:flare}. For clearer demonstration, we also plot the bolometric LCs of two sources (i.e., SN 2019stc and PS1-12cil) in Figure \ref{fig:bol}, along with the underlying energy injection and the flare component respectively. The bolometric LCs were constructed using \superbol \footnote{\href{https://github.com/mnicholl/superbol}{https://github.com/mnicholl/superbol}} \citep{Nic18} based on the observed fluxes in available filters. We simply fit their spectral energy distribution (SED) assuming blackbody radiation\footnote{As the lack of ultraviolet (UV) and near-infrared (NIR) data, extrapolations were invoked in the process of fitting.}. Time dilation and the Galactic extinction were corrected. We succeed in reproducing the bumpy bolometric LCs by performing the same fitting procedure adopted for multi-color LCs. The best parameters are consistent with those from the multi-color fitting.


\begin{table}[ht!]
\footnotesize
\caption{Free Parameters and Priors Used in Our Calculations}
\begin{center}
\begin{tabular*}{0.47\textwidth}{lll} \hline\hline
Parameter & Definition &Prior \\ 
\hline 
$M_{\rm ej}/M_\odot$ & Ejecta mass& [0.5, 50]\\
$P_{\rm i}/{\rm ms}$ & Initial magnetar spin period & [0.7, 20] \\
$\log_{10}(B_{\rm p}/10^{14}{\rm G})$ & Polar magnetic field & [-2, 1]\\ 
$v_{\rm ej}/10^4 {\rm km \ s}^{-1}$ & Ejecta velocity& [1, 3]\\ 
$\log_{10}(\kappa_{\rm \gamma}/{\rm cm}^2{\rm g}^{-1})$ & High energy opacity& [-3, 2]\\
$\kappa/{\rm cm}^2{\rm g}^{-1}$ & Optical opacity& [0.05, 0.34]\\ 
$T_{\rm floor}/10^3{\rm K}$ & Photosphere temperature floor& [3, 20] \\
$t_{\rm shift}/{\rm d}$ & Time interval between & [0.1, 30]\\
& explosion and the first detection& \\ 
$t_{\rm 0}/{\rm d}$ & Start time of flare injection& [30,150]\\ 
$t_{\rm flare,rise}$/${\rm d}$ & Flare rise time& [10, 100]\\ 
$L_{\rm flare,pk}/{\rm 10^{42} erg\,s^{-1}}$ & Flare luminosity& [0.1, 100]\\ 

\hline
\end{tabular*}
\label{tab:def}
\end{center}
\tablecomments{The last three parameters are highly dependent on the light curve features of the specific object, where $t_{\rm flare,rise}$ is the theoretical estimation rather than an observable. Therefore, the chosen prior ranges are general. }
\end{table}

\begin{table*}[ht!]
\begin{centering}
\footnotesize
\caption{\label{tab:basepara} Medians and $1\sigma$ Errors for Baseline Parameters}
\begin{tabular}{lcccccccc} \hline\hline
Name & $M_{\rm ej}$ &$P_{\rm i}$ & $B_{\rm p}$&$v_{\rm ej}$ & $\rm \kappa_{\gamma}$ & $\kappa$ & $T_{\rm floor}$ &$t_{\rm shift}$\\ 
& ($M_\odot$) & $({\rm ms})$ &$(10^{14}{\rm G})$ &$(10^4{\rm km \ s}^{-1})$ & $({\rm cm}^2{\rm g}^{-1})$ & $({\rm cm}^2{\rm g}^{-1})$ & $(10^3{\rm K})$& $({\rm d})$\\

\hline 
PS1-12cil & $2.71^{+1.34}_{-0.65}$& $2.57^{+0.31}_{-0.33}$&$3.09^{+0.21}_{-0.21}$&$0.92^{+0.02}_{-0.01}$ & $0.72^{+1,24}_{-0.18}$&$0.08^{+0.03}_{-0.02}$& $5.77^{+1.31}_{-1.83}$&$3.41^{+0.20}_{-0.19}$ \\
SN 2018kyt &$18.03^{+7.59}_{-6.24}$& $6.24^{+0.06}_{-0.07}$&$2.04^{+0.09}_{-0.09}$&$0.82^{+0.03}_{-0.03}$ & $3.55^{+9.07}_{-0.09}$&$0.03^{+0.02}_{-0.01}$& $6.66^{+0.17}_{-0.19}$&$12.98^{+0.57}_{-0.62}$ \\ 
SN 2019lsq & $2.22^{+0.54}_{-0.42}$& $3.52^{+0.11}_{-0.12}$&$0.83^{+0.11}_{-0.13}$&$0.75^{+0.01}_{-0.02}$ & $0.02^{+0.01}_{-0.01}$&$0.23^{+0.06}_{-0.05}$& $12.34^{+0.14}_{-0.13}$&$8.09^{+0.45}_{-0.31}$ \\
SN 2019stc & $4.49^{+2.13}_{-1.32}$& $5.68^{+0.27}_{-0.30}$&$1.26^{+0.20}_{-0.23}$&$0.84^{+0.03}_{-0.03}$ & $0.01^{+0.01}_{-0.01}$&$0.16^{+0.06}_{-0.05}$& $5.28^{+0.10}_{-0.09}$&$9.60^{+0.93}_{-0.73}$\\
SN 2021mkr & $16.80^{+5.93}_{-3.36}$& $1.73^{+0.12}_{-0.13}$&$0.25^{+0.06}_{-0.06}$&$3.18^{+0.56}_{-0.45}$ & $0.05^{+0.02}_{-0.02}$&$0.20^{+0.08}_{-0.05}$& $9.25^{+0.12}_{-0.10}$&$11.36^{+0.66}_{-0.69}$\\ 
\hline
\end{tabular}
\end{centering}
\end{table*}

\begin{figure*}[ht!]
\centering
    \includegraphics[width=1\textwidth]{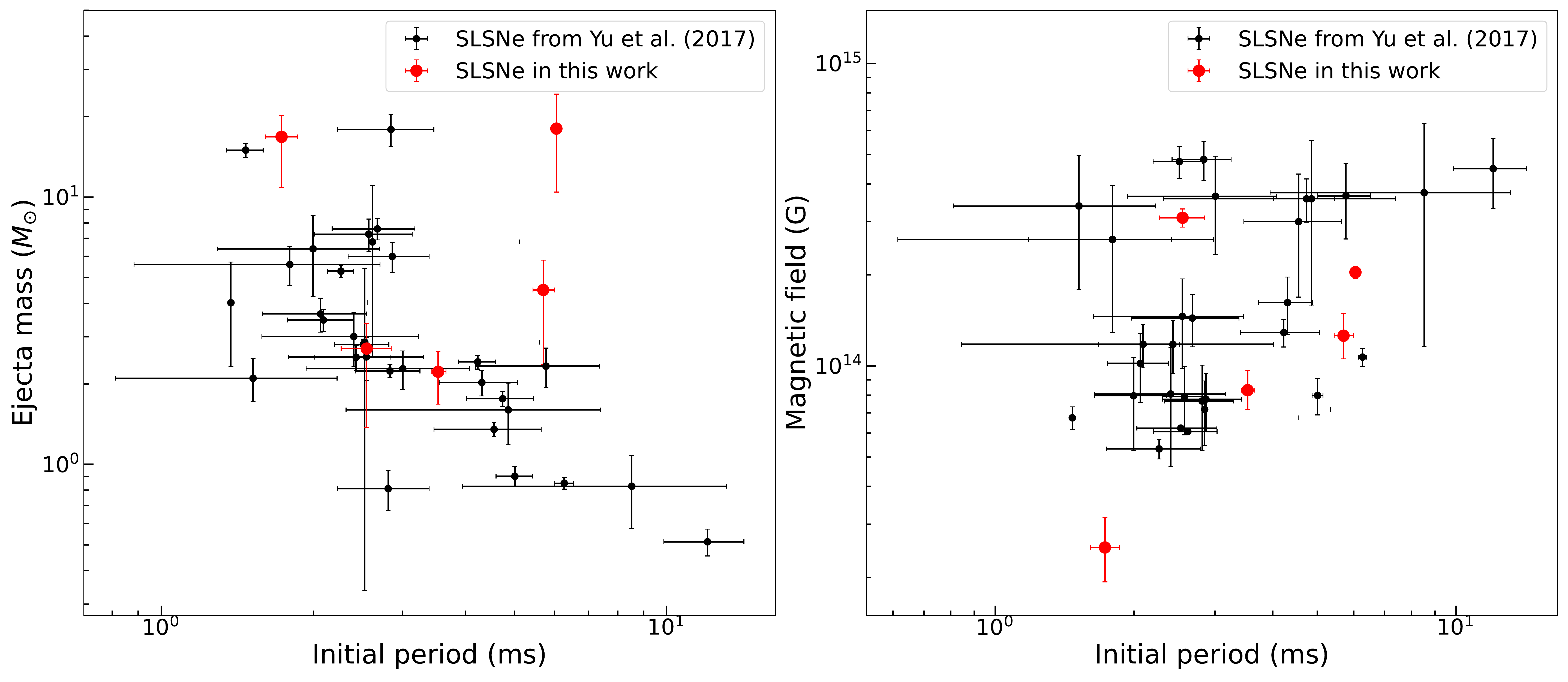}
    \caption{Comparison of baseline parameters (ejecta mass, initial period and magnetic field) of SLSNe-I in our sample (red dot) with those of SLSNe-I from \citet{Yu17} (black dot).}
    \label{fig:base_compare}
\end{figure*}
\section{Results} \label{sec:results}


In this section, we present the quantitative analysis of the fitting parameters obtained for the basic magnetar and flares, and also compare the flare properties with those of GRBs. The three key parameters for the underlying magnetar are $M_{\rm ej}$, $P_{\rm i}$ and $B_{\rm p}$. In our sample, the ejecta mass ranges in $(2.22-18.03)~M_{\odot}$, the initial period are distributed in $(1.73-6.24)$ ms, and the magnetic field scatter between $(0.25-3.09)\times 10^{14}\rm G$. We plot the key parameters of our sample in two parameter planes (the $M_{\rm ej}$ and $B_{\rm p}$ versus $P_{\rm i}$) in Figure \ref{fig:base_compare}, where 31 SLSNe-I are also shown for comparison. The baseline parameters of our sample roughly lie within the typical range of SLSNe-I, indicating that the basic energy injection of the undulated sources in our work has no specificity. 

In terms of flares, we extracted six flares from the 5 selected SLSNe-I. All the flares were injected $(3.4-7.4)\times 10^6 \operatorname{~s}$ after the explosion. The rise time distributes in a range of $(1.8-7.3)\times 10^6 \rm \operatorname{~s}$, and mainly within $(1.8-3.5)\times 10^6 \rm \operatorname{~s}$. Their peak luminosity ranges in $(0.85-43)\times 10^{43}\rm \operatorname{erg~s}^{-1}$. 

\begin{table}[ht!]
\caption{Medians and $1\sigma$ Errors for Flare Parameters}
\begin{center}
\begin{tabular}{lccc} \hline\hline
Name & $t_{0}$ &$t_{\rm flare,rise}$ & $ L_{\rm flare,pk}$\\ 
& ($\rm d$) & $({\rm d})$ &$(10^{42}\rm {erg \ s}^{-1})$\\
\hline 
PS1-12cil & $39.98^{+0.62}_{-0.66}$& $26.18^{+0.80}_{-0.75}$&$31.93^{+0.58}_{-0.57}$ \\
SN 2018kyt &$60.68^{+1.75}_{-1.82}$& $40.09^{+1.91}_{-1.78}$&$8.54^{+0.27}_{-0.26}$ \\ 
SN 2019lsq & $45.04^{+1.74}_{-1.74}$& $29.29^{+2.07}_{-2.04}$&$60.32^{+4.07}_{-3.83}$\\
&$41.16^{+4.00}_{-3.66}$ & $84.87^{+4.41}_{-4.45}$&$124.48^{+4.46}_{-4.24}$ \\
SN 2019stc & $85.81^{+1.81}_{-2.04}$& $38.88^{+2.01}_{-1.76}$&$39.93^{+1.10}_{-1.13}$\\ 
SN 2021mkr & $62.90^{+1.73}_{-1.89}$& $21.64^{+2.49}_{-2.18}$&$428.07^{+29.84}_{-29.40}$\\ 
\hline
\end{tabular}
\end{center}
\label{tab:flare}
\tablecomments{$t_{\rm flare,rise}$ represents the rise time of a flare luminosity from injection to the peak, which can be converted to the flare peak time ($t_{\rm pk}$) by adding the $t_{\rm 0}$. }
\end{table}

\begin{figure}[ht!]
\centering
    \includegraphics[width=0.5\textwidth]{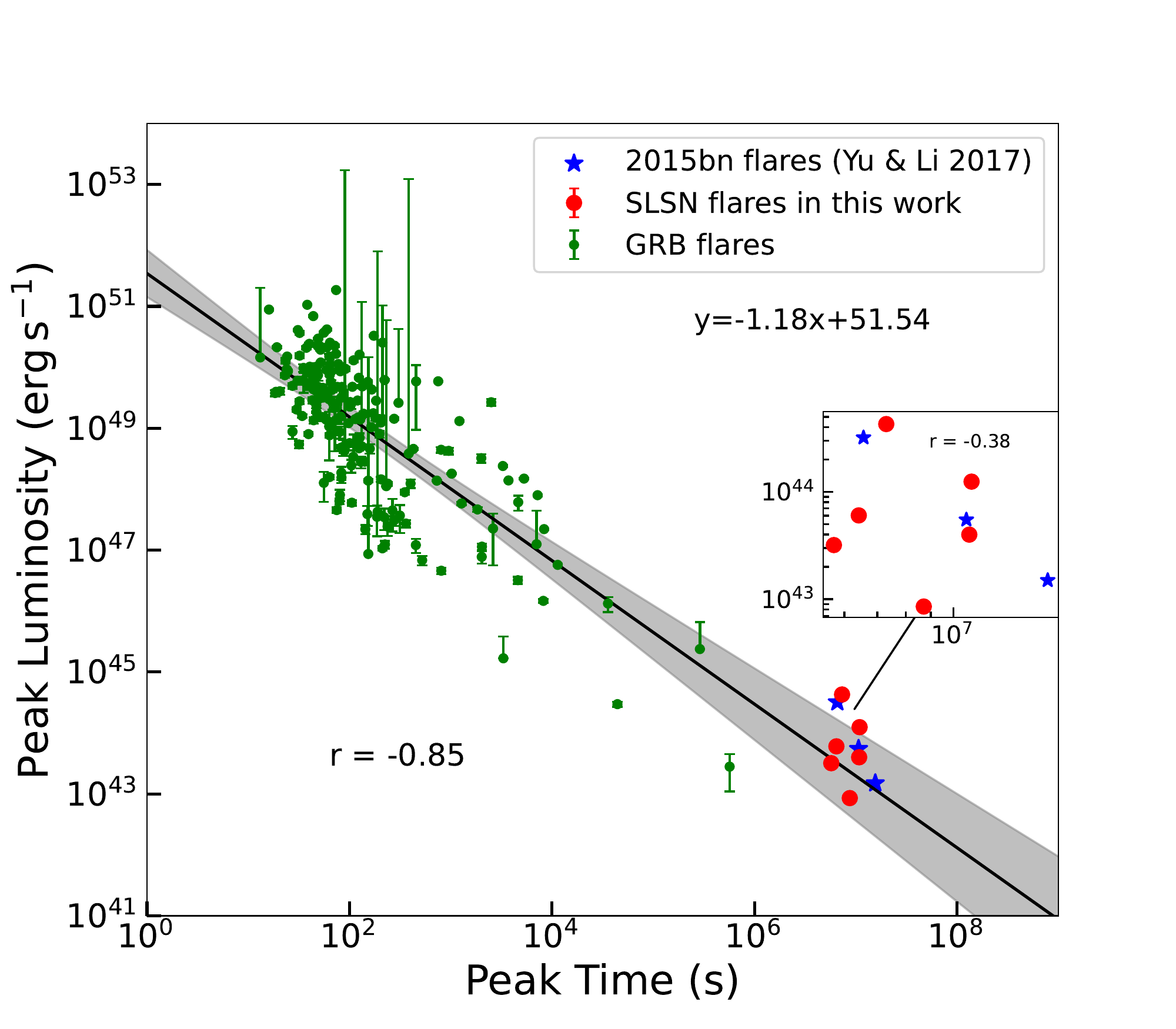}
    \caption{Comparison of SLSNe flares (red dot and blue star) with 200 GRB flares (green dot) in the peak luminosity versus peak time parameter space. The GRB data are obtained from \citet{yi2016}. Flare parameters of SN 2015bn (blue star) are taken from \citet{Yu&Li2017}. The solid black line represents the correlation between the peak luminosities and peak times taking both the GRB and SLSN flares into account. $3\sigma$ region of the fitting line is shown as the grey shadow. The small box on the right zooms in the SLSNe flares. The Pearson correlation coefficient ``r'' is shown.}
    \label{fig:flarerelation}
\end{figure}

In general, the SLSN flares in our sample last much longer and are much dimmer than the GRB flares (with the peak times and duration mainly between $100-1000~\rm s$ \citep{yi2016}).  We plot the flare parameters ($L_{\rm flare,pk}$ and $t_{\rm pk}$\footnote{$t_{\rm pk}=t_{\rm flare,rise}+t_{\rm 0}$, which reflects the peak time of the flare.}) of our sample with 200 GRB flares from \citet{yi2016} in Figure \ref{fig:flarerelation}. They found that the GRB flare parameters follows an empirical relation of 

\begin{equation}
     \log L_{\rm flare,pk}= 51.73-1.27\log t_{\rm pk}.
\end{equation}

The study conducted by \citet{Yu&Li2017} reveals that the three flares of SN 2015bn are situated within the 2$\sigma$ range of the relation established in the context of GRB flares (represented by blue stars in Figure \ref{fig:flarerelation}). A similar result was found in our work. After adding flares of our sample, we refit the GRB and SLSN flare parameters and found that the relation still stands, which can be expressed as 
\begin{equation}
    \log L_{\rm flare,pk}= 51.54-1.18\log t_{\rm pk}.
\end{equation}
This relation has a stronger Pearson coefficient reaching $-0.85$, suggesting that the more luminous the flare, the later it appears. However, we could not find such a relation in the SLSN sample only, which may be due to the small sample size.

\section{Conclusions and discussion} \label{sec:discussion}

We have presented the systematic analysis of a sample of SLSNe-I with late-time undulations under the central magnetar assumption. We employed a semi-analytical magnetar model with additional flare activities to fit their multi-color LCs, as well as two available bolometric LCs, and compared the flare properties with GRB flares. Our main conclusions are as follows;

\begin{enumerate}

\item The undulating LCs of the 5 selected SLSNe-I can be explained by additional flare activities from a central spin-down magnetar. On the one hand, the underlying LC profile of these objects is consistent with the magnetar-driven baseline, and the corresponding parameters ($P_{\rm i}$, $M_{\rm ej}$ and $B_{\rm p}$) fall within the typical range observed for SLSNe-I (see Figure \ref{fig:base_compare}), suggesting no peculiar features existing in their basic energy sources. On the other hand, the distinct late-time bumps in their multi-color LCs can be accurately replicated by invoking magnetar flares. 

\item Six flares were extracted from the LCs of the 5 selected SLSNe-I. The injection of flares started between day 39 and day 86 after the explosion, and it took 21 to 84 days to reach the peak luminosity, which ranged in $(0.85-43)\times 10^{43}\rm \operatorname{erg~s}^{-1}$. 

\item SLSN flares tend to be systematically dimmer and longer than GRB flares. The strong correlation between peak luminosity and peak time in GRB flares remains evident and is even stronger ($r=-0.85$) when including the SLSNe-I sample. This is reasonable given the intrinsically weaker magnetic field in SLSNe-I compared to GRBs. However, we do not find a similar correlation in the SLSN sample alone, possibly due to the small sample size. 
\end{enumerate}

In previous literature, different models have been discussed to explain the late-time undulations. Within the framework of magnetar models, \cite{chugai2022} proposed that bumps are caused by the magnetar dipole field enhancement several months after the explosion. \cite{Moriya2022} suggested that the light-curve bumps are caused by variations of the thermal energy injection from magnetar spin down. However, \cite{chugai2022} only provide an explanation for a single bump and \cite{Moriya2022} have predicted an increase in photospheric temperature which was not observed in SN 2020qlb. 
In addition to the magnetar model, other models also include the interaction model \citep[e.g.,][]{Inserra2017Interaction,Fiore2017gci,LiCilCSM,SN2020qlb}, the sudden drop of the ejecta opacity \citep{MetzgerOpacitydrop}, neutron stars colliding with binary companions \citep{Hirai2022}. Undulations may also be caused by an asymmetry in the geometry of the ejected material \citep{Kaplan2020}. In this paper, we believe that the late-stage flare activity of the central magnetar can well explain the undulation in the LC. 

In general, the magnetar assumption accommodates the systematic diversity of the massive progenitors, allowing for a wide range of magnetic fields to exist in magnetars with similar behaviors. This aligns with the correlation between late-time undulations of SLSNe-I and GRB flares in the $L_{\rm flare,pk}$ verses $t_{\rm pk}$ plane, providing supporting evidence that they share a common power origin. The linearity between peak time and luminosity over many orders of magnitude demonstrates that a dimmer flare is likely to peak at a later time, with a consistent radiated energy of $\sim 10^{51}~\rm erg$. This suggests that there may be regular pulsive activities of these newborn magnetars. If this is the case, the flares of SLSNe-I may also be produced through magnetic reconnection processes similar to solar flares \citep{ShibataMHD}, resulting in shock heating, particle acceleration, and energy ejection. 

Although we have drawn conclusions, it is important to note that our findings are based on a limited sample size. To gain further insight into the power source of SLSNe and investigate any potential relation between SLSN and GRB, more observations of well-sampled undulated phases are necessary. Fortunately, upcoming telescopes and surveys such as Legacy Survey of Space and Time (LSST), ZTF phase II, and James Webb Space Telescope (JWST) will likely discover numerous SLSNe with clearly defined light curves in the near future. This will allow for a more comprehensive understanding of their explosion mechanisms and corresponding progenitors.

\begin{acknowledgments}
We thank Zongkai Peng and Zhihao Chen for helpful comments and discussions, and an anonymous referee for important suggestions. This work is supported by the National Natural Science Foundation of China (Projects 12021003), the National SKA Program of China (2022SKA0130100), and the National Key R$\&$D Program of China (2021YFA0718500). The ZTF forced-photometry service was funded under the Heising-Simons Foundation grant \#12540303 (PI: Graham).
\end{acknowledgments}

\bibliography{SLSNe-I}{}
\bibliographystyle{aasjournal}

\end{CJK*}
\end{document}